\documentclass{article}
\newcommand{\lev}{{*\!F}_{\alpha\beta\mu\nu}}
\newcommand{\pr}{{F\!*}_{\alpha\beta\mu\nu}}
\newcommand{\dv}{{*\!F\!*}_{\alpha\beta}{}^{\mu\nu}}
\newcommand{\ef}{F_{\alpha\beta\mu\nu}}
\newcommand{\es}{\mathcal{S}_{\alpha\beta\mu\nu}}
\newcommand{\err}{\mathcal{R}_{\alpha\beta\mu\nu}}
\newcommand{\dva}{{*\!F\!*}_{\alpha\beta\mu\nu}}
\begin{document}

\title{Nonabelian duality and analytic continuation of instantons }%
\author{A. L. Koshkarov\thanks{email: Koshkarov@petrsu.ru}}%
\date{University of Petrozavodsk, Physics department, Russia}
\maketitle
\begin{abstract}
Within the framework of the gauge $O(1,3)\times O(1,3)$-theory, an
extension of the Belavin-Polyakov-Schwarz-Tyupkin ansatz  is
proposed by incorporation there the Levi-Civita tensor. The
duality properties of the theory, admitting introduction the
complex structure, are such that selfduality condition of the
field tensor is an equation for complex-analytic function. New
type of duality is found out.
\end{abstract}
\section{Introduction}

The instanton theory, having started with the paper~\cite{bpst}, looks quite
completed. Instantons, which seem to describe quantum vacuum transitions,
in the {\em Euclidean} theory possess nice topological characteristics.

Such a visible geometric treatise is absent in the physical pseudoeuclidean space-time.
Still close relation between the Euclidean instantons and selfdual solutions
in the pseudoeuclidean space-time  may well exist.

To make it clear we treat nonabelian gauge theory with the gauge
group $O(1,3)$ in the physical 4-dimensional pseudoeuclidean
space-time. It is the Lorenz group $O(1,3)$ that is the part of
the motion group. It is significant for us that {\em real} group
$O(1,3)$ is related to {\em complex} groups $L(C)$ and $SL(2,C)$.

Of course on one hand, self-dual solutions can be obtained from the
$O(4)$-instantons just by the Euclid rotation. However, we try to show that
the pseudoeuclidean spacetime treatment is far more general for $O(1,3)$-self-dual
solutions are {\em complex-analytic} manifolds  hence they include as a special
case Euclidean solutions.

Essential point in our approach is that gauge group is a part of the motion
group. It permits  all indices (regardless internal or  the spacetime one)
to treat uniformly. This makes the theory simpler in the end. For example,
well-known the t'Hooft symbols~\cite{ho} $\eta^a_{\mu\nu}$ in the simple way
 express by Minkowski space tensors~\cite{nevaku}. Further we are going to
 apply these ideas to gravity since it can be treated as $O(1,3)$-gauge theory
with the group action in tangent space.

The approach appears to relate to new kind of {\em duality}. It
means as follows. Initial theory operates on fields
$A_\mu=(1/2)FA_{\alpha\beta\mu} T^{\alpha\beta}$ and
$F_{\mu\nu}=(1/2)F_{\alpha\beta\mu\nu}T^{\alpha\beta}$, where
$T^{\alpha\beta}$ ---the internal symmetry group generators, and
the field tensor obeys the Bianchi identity:
$D^\nu{F\!*}_{\mu\nu}\equiv0$. One can consider a field
$\tilde{A}_{\alpha}= 1/2\tilde{A}_{\alpha\mu\nu}J^{\mu\nu}$ to be
"dual"  to starting field ($J^{\mu\nu}$---the space-time Lorenz
group generators). The field
$\tilde{F}_{\alpha\beta}=(1/2)\tilde{F}_{\alpha\beta\mu\nu}T^{\alpha\beta}$
corresponds to $\tilde{A}_{\alpha}$. There should be fulfilled an
identity $D^\beta{*\!\tilde{F}}_{\alpha\beta}\equiv0$ for this
field. It is the case in gravity.

\section{Extended Belavin-Polyakov-Schwarz-Tyupkin's ansatz}

Let us consider a nonabelian gauge field in {\em pseudoeuclidean} spacetime
$R_1^4$ with the group $O(1,3)$. The field is described  in routine way with
giving a potential $A_\mu(x)$ and calculating field strengths $F_{\mu\nu}(x)$
to take values in the Lie algebra of the Lorenz group:
$$
A_\mu=\sum_{\alpha<\beta}^{}A_{\alpha\beta\mu}T^{\alpha\beta}=
\frac{1}{2}A_{\alpha\beta\mu}T^{\alpha\beta},\quad
F_{\mu\nu}=\sum_{\alpha<\beta}^{}F_{\alpha\beta\mu\nu}T^{\alpha\beta}=
\frac{1}{2}F_{\alpha\beta\mu\nu}T^{\alpha\beta}
$$
Here $T^{\alpha\beta},\alpha<\beta$ --- 6 generators of $so(1,3)
\simeq su(2)\oplus su(2)$-algebra, which can be considered antisymmetric in
$\alpha,\beta$ for calculations convenience. The generators algebra is:
$$
[T^{\alpha\beta},T^{\gamma\delta}]=\frac{1}{2}(\varepsilon^{\alpha\beta\rho
\lambda}\varepsilon^{\gamma\delta\sigma}{}_\lambda-
\varepsilon^{\alpha\beta\sigma\lambda}\varepsilon^{\gamma\delta\rho}
{}_\lambda)T_{\rho\sigma}=
-\delta^{\alpha\gamma}T^{\beta\delta}-\delta^{\beta\delta}T^{\alpha\gamma}
+\delta^{\alpha\delta}T^{\beta\gamma}+\delta^{\beta\gamma}T^{\alpha\delta}
$$
Here the rotation generators $T^{ab}$ are anti-Hermitian. For convenience,
the Minkowski tensor denotes $\delta_{\mu\nu}$ or $\delta^{\mu\nu}$.

In the classical paper~\cite{bpst} (BPST), in order to find an explicit
solution in the Euclidean $O(4)$-gauge theory, $O(4)$-symmetric ansatz was
introduced:
\begin{equation}\label{1}
A_{\alpha\beta\mu}=f(r)x_\lambda\delta_{\alpha\beta\mu\lambda},\quad
\delta_{\alpha\beta\mu\lambda}=\delta_{\alpha\mu}\delta_{\beta\lambda}-
\delta_{\alpha\lambda}\delta_{\beta\mu}.
\end{equation}
This was the most general central-symmetric representation for
$O(4)$-gauge potential $A_\mu$, to have mixed the internal and space-time
indices.

Here more general ansatz is proposed:
\begin{equation}\label{2}
A_{\alpha\beta\mu}=x^\lambda(f(r)\delta_{\alpha\beta\mu\lambda}+g(r)
\varepsilon_{\alpha\beta\mu\lambda}),
\end{equation}
and gauge group is a Lorenz that $O(1,3)$ to be considered as an {\em internal}
symmetry group. Such an approach had already been treated in~\cite{nevaku}.

Motivations of the ansatz extension are quite clear. Technically the use of
fully antisymmetric tensor $\varepsilon_{\alpha\beta\mu\lambda}$ together with
or instead of $\delta_{\alpha\beta\mu\lambda}$ for representation of the
potential $A_{\alpha\beta\mu}$ is quite acceptable and not worse. Additionally
(\ref{2}) gives an opportunity to find a new solution, distinctive feature
of that will be possible asymmetry under reflections.

What extent are functions $u(r)$ and $v(r)$ independent in?

We'll see that these functions  are  parts of a single whole complex-analytic
function, if strength field tensor is selfdual.

Now we pass from the real potential $A_{\alpha\beta\mu}$ of the gauge Lorenz
group on to complex potential $a_{\alpha\beta\mu}$ of $SL(2,C)\times SL(2,C)$-group
(or {\em $L(C)$}):
\begin{equation}\label{3}
a_{\alpha\beta\mu}=A_{\alpha\beta\mu}+i{*\!A}_{\alpha\beta\mu},\quad
{*\!A}_{\alpha\beta\mu}=
\frac{1}{2}\varepsilon_{\alpha\beta\rho\sigma}A^{\rho\sigma}{}_\mu
\end{equation}
Then with taking into account~(\ref{2}):
\begin{equation}\label{4}
a_{\alpha\beta\mu}=(f-ig)x^\lambda(\delta_{\alpha\beta\mu\lambda}+
i\varepsilon_{\alpha\beta\mu\lambda})
\end{equation}
By now one can be seen that the potential of $SL(2,C)\times SL(2,C)$-group
(hence an entire theory) contains complex-analytic function $f-ig$. We can
treat the functions to depend on four variables $x^\mu$. Then further one can
 say of holomorphic continuation of solutions.

It is a {\em tensors}  of  the $SL(2,C)$-group appeared here that are
t'Hooft's symbols in the space $R_1^4$:
$$
\delta_{0a\mu\nu}
+ i\varepsilon_{oa\mu\nu}=\eta_{a\mu\nu}
$$
Later we are going to consider tensors
$\eta_{\alpha\beta\mu\nu}=\delta_{\alpha\beta\mu\nu}
+ i\varepsilon_{\alpha\beta\mu\nu}$, main property of which is double sided
(anti)selfduality (see below):
$$
{*\!\eta}_{\alpha\beta\mu\nu}={\eta\!*}_{\alpha\beta\mu\nu}=-i\eta_{\alpha\beta\mu\nu}
$$

We'll be back to this later, and for now go on to work out the ansatz~(\ref{2}).
Our actual aim is to calculate (real) field strength tensor. But first let us
usher in some notations  and compose a list of formulas to require next.

First of all, we introduce {\em dual conjugation} operations. For any pair
of indices $\alpha,\beta$, in which a tensor  is antisymmetric, we define
a {\em dual} tensor. For instance, gauge $o(1,3)$-potential $A_{\alpha\beta\mu}$
is antisymmetric in $\alpha,\beta$. Dual tensor ${*\!A}_{\alpha\beta\mu}$ is
$$
{*\!A}_{\alpha\beta\mu}=\frac{1}{2}\varepsilon_{\alpha\beta\rho\sigma}
A^{\rho\sigma}{}_\mu
$$

For strength tensor $F_{\alpha\beta\mu\nu}$ (and for others) it
defines the left dual conjugate ($\lev$), the right dual conjugate
($\pr$), twice
 dual conjugate ($\dva$):
$$
\lev=\frac{1}{2}\varepsilon_{\alpha\beta\rho\sigma}F^{\rho\sigma}{}_{\mu\nu},\:
 \pr=\frac{1}{2}F_{\alpha\beta}{}^{\rho\sigma}\varepsilon_{\rho\sigma\mu\nu},\:
\dv=\frac{1}{2}\varepsilon_{\alpha\beta\rho\sigma}
F^{\rho\sigma}{}_{\gamma\delta}\varepsilon^{\gamma\delta\mu\nu}
$$
E. g.
$$
{*\!\delta}_{\alpha\beta\mu\nu}={\delta\!*}_{\alpha\beta\mu\nu}=
\varepsilon_{\alpha\beta\mu\nu},\, {*\!\varepsilon}_{\alpha\beta\mu\nu}=
{\varepsilon\!*}_{\alpha\beta\mu\nu}=-\delta_{\alpha\beta\mu\nu},
$$
\begin{equation}\label{5}
{*\!\delta\!*}_{\alpha\beta\mu\nu}=-\delta_{\alpha\beta\mu\nu},\,
{*\varepsilon\!*}_{\alpha\beta\mu\nu}=-\varepsilon_{\alpha\beta\mu\nu}
\end{equation}
Double left (right) dual conjugation operation gives:
\begin{equation}\label{6}
*\!{*\!F}_{\alpha\beta\mu\nu}=-\ef={{F\!*}\!*}_{\alpha\beta\mu\nu}
\end{equation}
Twice dual tensor $\dv$ can be expressed by $\ef$ and its contractions~\cite{ko1}:
\begin{equation}\label{7}
\dva=-F_{\mu\nu\alpha\beta}+F_{\alpha\mu}\delta_{\beta\nu}+F_{\beta\nu}
\delta_{\alpha\mu}-F_{\alpha\nu}\delta_{\beta\mu}-
F_{\beta\mu}\delta_{\alpha\nu}-\frac{1}{2}F\delta_{\alpha\beta\mu\nu}
\end{equation}
Here
$$
F_{\alpha\mu}=F_\alpha{}^\nu{}_{\mu\nu},\quad F=F^\mu{}_\mu
$$

Professor D. Fairlie kindly informed me of the remarkable Lanczos
paper~\cite{la}, where betweenness relation $R_{\alpha\beta\mu\nu}$ and
${*\!R\!*}_{\alpha\beta\mu\nu}$ probably first had been obtained for the Euclidean signature space-time.

Further tensor
\begin{equation}\label{8}
X_{\alpha\beta\mu\nu}=x_\alpha x_\mu\delta_{\beta\nu}+x_\beta x_\nu
\delta_{\alpha\mu}-x_\alpha x_\nu\delta_{\beta\mu}-x_\beta x_\mu\delta_{\alpha\nu}
\end{equation}
often occurs. Here is  some properties of the tensor:
\begin{equation}\label{9}
{*\!X}_{\alpha\beta\mu\nu}=(x_\nu\varepsilon_{\alpha\beta\mu\lambda}-
x_\mu\varepsilon_{\alpha\beta\nu\lambda})x^\lambda,\quad
{X\!*}_{\alpha\beta\mu\nu}=(x_\beta\varepsilon_{\alpha\mu\nu\lambda}-
x_\alpha\varepsilon_{\beta\mu\nu\lambda})x^\lambda,
\end{equation}
$$
{*\!X\!*}_{\alpha\beta\mu\nu}=X_{\alpha\beta\mu\nu}-r^2
\delta_{\alpha\beta\mu\nu},\;{*\!X}_{\alpha\beta\mu\nu}+
{X\!*}_{\alpha\beta\mu\nu}=r^2\varepsilon_{\alpha\beta\mu\nu},
$$
$$
X_{\alpha\mu}=X_\alpha{}^\nu{}_{\mu\nu}=r^2\delta_{\alpha\mu}+
2x_\alpha x_\mu,\quad
 X=X^\mu{}_\mu=6r^2
$$

Now we begin to calculate the strength tensor of $o(1,3)$-field $\ef$:
$$
\ef=\partial_\mu A_{\alpha\beta\nu}-\partial_\nu A_{\alpha\beta\mu}+
[A_\mu,A_\nu]_{\alpha\beta}
$$
Using~(\ref{9}), it  directly obtains:
$$
\partial_\mu A_{\alpha\beta\nu}-\partial_\nu A_{\alpha\beta\mu}=
-2(f\delta_{\alpha\beta\mu\nu}+g\,\varepsilon_{\alpha\beta\mu\nu})-\frac{1}{r}
(f'X_{\alpha\beta\mu\nu}+g'{*\!X}_{\alpha\beta\mu\nu})
$$
The commutator calculation require a little more complicated
computations. Using the generators algebra and
formulas~(\ref{2})---(\ref{9}), we find:
$$
[A_\mu,A_\nu]_{\alpha\beta}=A_{\lambda\beta\mu}A^\lambda{}_{\alpha\nu}-
A_{\lambda\alpha\mu}A^\lambda{}_{\beta\nu}=(f^2-g^2)(X_{\alpha\beta\mu\nu}
-r^2\delta_{\alpha\beta\mu\nu})-2fg{X\!*}_{\alpha\beta\mu\nu}
$$
At last, the strength tensor is obtained:
$$
\ef=(-2f-r^2(f^2-g^2))\delta_{\alpha\beta\mu\nu}-
(2g+rg')\varepsilon_{\alpha\beta\mu\nu}+
(f^2-g^2-\frac{f'}{r})X_{\alpha\beta\mu\nu}-
$$
\begin{equation}\label{10}
(2fg-\frac{g'}{r}){X\!*}_{\alpha\beta\mu\nu}
\end{equation}

So, the field $\ef$ is represented in the form of decomposition in
tensors $\delta_{\alpha\beta\mu\nu}$, $X_{\alpha\beta\mu\nu}$ and
their dual conjugates. This decomposition is rather convenient in
order to study the field properties. For example, by means of the
use~(\ref{5}---\ref{9}) properties, one can easily find different
dual conjugates.

The presence of the tensor $\varepsilon_{\alpha\beta\mu\nu}$ in
decomposition~(\ref{10}) suggests that $T$- and $P$-symmetries are not
conserved for the field~(\ref{10}). However, for $sl(2,C)$-field (see below)
these symmetries are not broken.

Note, that for $g=0$ a property
$$
\ef=F_{\mu\nu\alpha\beta}
$$
takes place like in gravity. But it is not the case for~(\ref{10}) because of
the last term ${X\!*}_{\alpha\beta\mu\nu}$.

The Bianchi identities are of the form:
\begin{equation}\label{toj1}
D_\nu {F\!*}_{\alpha\beta\mu}{}^\nu\equiv0,
\end{equation}
or as follows:
\begin{equation}\label{toj2}
D_\nu {*\!{F\!*}}_{\alpha\beta\mu}{}^\nu\equiv0,
\end{equation}

Now we pay attention to a new type of nonabelian duality to appear here. High
symmetry of the ansatz~(\ref{2}) allows to operate on {\em dual} potential
$\tilde{A}_{\alpha}=(1/2)\tilde{A}_{\alpha\mu\nu}J^{\mu\nu}$ with
correspondent dual strength $\tilde{F}_{\alpha\beta}=
(1/2)\tilde{F}_{\alpha\beta\mu\nu}T^{\alpha\beta}$. Like~(\ref{10}), this
field can be formally calculated, not asking (as yet) a question
about physical meaning
of coordinates $x^\alpha$ of the space to act the internal symmetry group:
$$
\tilde{F}_{\alpha\beta\mu\nu}=(-2f-r^2(f^2-g^2))\delta_{\alpha\beta\mu\nu}-
(2g+rg')\varepsilon_{\alpha\beta\mu\nu}+
(f^2-g^2-\frac{f'}{r})X_{\alpha\beta\mu\nu}-
$$
$$
(2fg-\frac{g'}{r}){*\!X}_{\alpha\beta\mu\nu}
$$
This almost coincide with~(\ref{10}) (the difference is in the last term:
${*\!X}$ instead of ${X\!*}$):
$$
\ef=\tilde{F}_{\alpha\beta\mu\nu}+(2fg-\frac{g'}{r})({*\!X}_
{\alpha\beta\mu\nu}-{X\!*}_{\alpha\beta\mu\nu})
$$

This field should obey to Bianchi's identity, which here is of the form:
$$
\tilde{D}^\beta{*\!\tilde{F}}_{\alpha\beta\mu\nu}\equiv0,\quad
\tilde{D}^\beta=\partial^\beta+[\tilde{A}^\beta,]
$$
From this and from the relation between $F$ and $\tilde{F}$ it follows one
more identity on the field $F$:
$$
\tilde{D}^\beta ({\!*F}_{\alpha\beta\mu\nu}+(\frac{g'}{r}-2fg)({*\!X}_
{\alpha\beta\mu\nu}-{X\!*}_{\alpha\beta\mu\nu}))\equiv0
$$
and as well:
$$
\tilde{D}^\beta ({\!*F\!*}_{\alpha\beta\mu\nu}+(\frac{g'}{r}-2fg)({*\!X\!*}_
{\alpha\beta\mu\nu}+X_{\alpha\beta\mu\nu}))\equiv0
$$

Let us be back to the field $\ef$. Equations
\begin{equation}\label{11}
\ef=\pm\dva
\end{equation}
are important.

If the equations~(\ref{11}) hold, then on account of the Bianchi
identity the Yang-Mills equations for the field $\ef$ will hold
too. Therefore, on the base of~(\ref{11}) it is convenient to
introduce~\cite{ko1} tensors $\es$ and $\err$:
$$
\es=\frac{1}{2}(\ef+\dva),\quad  \err=\frac{1}{2}(\ef-\dva)
$$
Then now (\ref{11}) are equivalent to equations:
\begin{equation}\label{12}
\es=0,
\end{equation}
\begin{equation}\label{13}
\err=0
\end{equation}
These equations are equivalent to the (anti)selfduality conditions.
It is true for groups $SL(2,C)(SU(2))$ as well.

The application of these equations to gravity enabled to generalize and
extend General Relativity~\cite{kosh}. The gravitational analog of~(\ref{12})
is an inhomogeneous equation and includes an energy-momentum current.
It would be not bad to learn something about the structure of the Yang-Mills
current taking into account the gravitational analogy and using some
phenomenological arguments\footnote{Cosmas Zachos
pointed out that expressing  covariant derivative of the curvature tensor
by  the energy-momentum current can be obtained without references to
duality just by contracting the Bianchi identity, then
taking into account Einstein's equation.}

The tensors $\es$ can be calculated with using the decomposition~(\ref{10})
and the properties~(\ref{5})---(\ref{9}):
$$
2\es=(-r^2f^2+r^2g^2+rf')(\delta_{\alpha\beta\mu\nu}-\frac{2}{r^2}X_{\alpha\beta\mu\nu})+
(2r^2fg-rg')(\varepsilon_{\alpha\beta\mu\nu}-\frac{2}{r^2}{X\!*}_{\alpha\beta\mu\nu})
$$
Then the equation~(\ref{12}) reduces to
$$
f'=r(f^2-g^2),\qquad
g'=2rfg
$$
Multiplying  the second formula by imaginary unit, then plus or minus the
first one, we obtain:
\begin{equation}\label{14}
f'\pm ig'=r(f\pm ig)^2
\end{equation}
Similarly, (\ref{13}) gives
$$
2\err=-(4f+rf'+r^2(f^2-g^2))\delta_{\alpha\beta\mu\nu}-
(4g+rg'+2r^2fg)\varepsilon_{\alpha\beta\mu\nu}=0
$$
from here
$$
4f+rf'+r^2(f^2-g^2)=0,\quad 4g+rg'+2r^2fg=0
$$
Again, multiplying  the second equation by $i$, then plus or minus the first
formula, we have
\begin{equation}\label{15}
4(f\pm ig)+r(f'\pm ig')+r^2(f\pm ig)^2=0
\end{equation}

So real conditions of (anti)selfduality of the gauge $O(1,3)$-theory
within the frame of~(\ref{2}) with real functions $f$ and $g$ are equivalent
to the equations~(\ref{14}),(\ref{15}) for complex-analytic function!

This result can be easier obtained by going over to $SL(2,C)\times SL(2,C)$-strength:
$$
\frac{1}{2}(\ef\pm i\lev)=\left(-(f\mp ig)-r^2(f\mp ig)^2\right)(\delta_{\alpha\beta\mu\nu}
\pm i\varepsilon_{\alpha\beta\mu\nu})+
$$
$$
\frac{1}{2}\left((f\mp ig)^2-\frac{1}{r}(f'\mp
ig')\right)(X_{\alpha\beta\mu\nu}\pm i\!*X_{\alpha\beta\mu\nu})
$$
And most direct way to obtain an equation for function $f\pm ig$
as a selfduality condition is from the very beginning, on the base~(\ref{4})
to use complex $sl(2,C)$-potential:
\begin{equation}\label{16}
a_\mu=\frac{1}{2}a_{\alpha\beta\mu}t^{\alpha\beta}=\frac{1}{2}u(\zeta)
(\delta_{\alpha\beta\mu\lambda}+i\varepsilon_{\alpha\beta\mu\lambda})
x^\lambda t^{\alpha\beta}
\end{equation}
where
$r\rightarrow \zeta$ (and $x^\mu$) --- are complex, and
$$
t^{\alpha\beta}=\frac{1}{2}(T^{\alpha\beta}+i{*\!T}^{\alpha\beta}),\quad
[t^{\alpha\beta},t^{\gamma\delta}]=-\delta^{\alpha\gamma}t^{\beta\delta}-
\delta^{\beta\delta}t^{\alpha\gamma}+\delta^{\alpha\delta}t^{\beta\gamma}+
\delta^{\beta\gamma}t^{\alpha\delta}
$$
To calculate this algebra is helpful to apply a formula:
$$
\varepsilon^{\alpha\beta\gamma\lambda}T^\delta{}_\lambda=\delta^{
\delta\alpha}{*\!T}^{\beta\gamma}+
\delta^{\delta\beta}{*\!T}^{\gamma\alpha}+\delta^{\delta\gamma}{*\!T}^{
\alpha\beta}
$$

Now the strengths are easily calculated:
$$
f_{\mu\nu}=\partial_\mu a_\nu-\partial_\nu a_\mu+[a_\mu,a_\nu]=\frac{1}{2}
t^{\alpha\beta}\left((\delta_{\alpha\beta\mu\nu}+i
\varepsilon_{\alpha\beta\mu\nu})(-u-\frac{1}{2}\zeta^2u^2)+\right.
$$
$$
\left.(X_{\alpha\beta\mu\nu}+i{*\!X}_{\alpha\beta\mu\nu})\frac{1}{2}(-\frac{u'}{\zeta}+u^2)\right)
$$

Thus  the field $f_{\mu\nu}$ is selfdual and complex-analytic manifold, if:
$$
\frac{1}{r}u'(\zeta)-u^2(\zeta)=0.
$$

\section{Other ans\"atze}

Complexifying by means of introduction into ansatz the Levi-Civita tensor can
be carried out for other substitutions as well. We demonstrate it  for
Corrigan-Fairlie-t'Hooft-Wilczek ansatz~\cite{fer} ($O(1,3)$-gauge group):
\begin{equation}\label{fair}
A_\mu=\frac{1}{2}A_{\alpha\beta\mu}T^{\alpha\beta},\quad A_{\alpha\beta\mu}=
\partial^\lambda\ln\phi(x)\,\delta_{\alpha\beta\mu\lambda}.
\end{equation}
It leads to the field
\begin{equation}\label{ef1}
F_{\alpha\beta\mu\nu}=\frac{2}{\phi^2}\phi_{\alpha\beta\mu\nu}-\frac{1}{\phi}
\varphi_{\alpha\beta\mu\nu}-\frac{\phi_\lambda\phi^\lambda}{\phi^2}
\delta_{\alpha\beta\mu\nu}
\end{equation}
Here
$$
\phi_\lambda=\partial_\lambda\phi,\quad \phi_{\alpha\beta}=\partial^2_
{\alpha\beta}\phi,\quad \phi_{\alpha\beta\mu\nu}=\phi_\alpha\phi_\mu
\delta_{\beta\nu}+\phi_\beta\phi_\nu\delta_{\alpha\mu}-\phi_\alpha\phi_\nu
\delta_{\beta\mu}-\phi_\beta\phi_\mu\delta_{\alpha\nu},
$$
$$
\varphi_{\alpha\beta\mu\nu}=\phi_{\alpha\mu}
\delta_{\beta\nu}+\phi_{\beta\nu}\delta_{\alpha\mu}-\phi_{\alpha\nu}
\delta_{\beta\mu}-\phi_{\beta\mu}\delta_{\alpha\nu}
$$
Conditions of selfduality~(\ref{12}) and antiselfduality~(\ref{13}) are,
respectively, of the form:
\begin{equation}\label{es1}
S_{\alpha\beta\mu\nu}=\left(\frac{1}{2\phi}\partial^\rho{}_\rho\phi-\frac{\phi_\lambda\phi^\lambda}
{\phi^2}\right)\delta_{\alpha\beta\mu\nu}+\frac{2}{\phi^2}\phi_{\alpha\beta\mu\nu}-
\frac{1}{\phi}\varphi_{\alpha\beta\mu\nu}=0
\end{equation}
\begin{equation}\label{er1}
\err=-\frac{1}{2}\frac{\partial^\rho{}_\rho\phi}{\phi}\;\delta_{\alpha\beta\mu\nu}=0,\quad
\end{equation}
We extend this ansatz:
\begin{equation}\label{fk}
A_{\alpha\beta\mu}=\partial^\lambda[\ln\phi(x)\,\delta_{\alpha\beta\mu\lambda}
+\ln\psi(x)\,\varepsilon_{\alpha\beta\mu\lambda}]
\end{equation}
$sl(2,c)\times sl(2,c)$-potential is
$$
A_{\alpha\beta\mu}+i{*\!A}_{\alpha\beta\mu}=\partial^\lambda
(\ln\phi-i\ln\psi)(\delta_{\alpha\beta\mu\lambda}
+i\varepsilon_{\alpha\beta\mu\lambda})
$$
Now we can see that e.g. antiselfduality condition~(\ref{er1}) is determined
by complex function $\Phi(x)$
$$
\ln\Phi=\ln\phi-i\ln\psi=\ln(\phi\exp(-i\ln\psi)),\quad \Phi=
\phi\exp(-i\ln\psi),
$$
which obeyed d'Alambert equation:
\begin{equation}\label{eq}
\partial^\rho{}_\rho\Phi=0
\end{equation}

One can  continue holomorphically the function $\Phi$ in variables
$x^\mu$, then take the complex t'Hooft~\cite{hof} solution to
equation~(\ref{eq}) with real $x^\mu$:
$$
\phi\exp(-i\ln\psi)=1+\frac{\lambda_1+i\kappa_1}{(x-x_1)^2}+\frac{\lambda_2+
\kappa_2}{(x-x_2)^2}+\ldots
$$
Separating here  real and imaginary parts, we find real $6q$-parametric
generalization of the $6q$-instantonic t'Hooft solution.

\section{Conclusion}

So in some cases, we managed to find some new solutions and continue analytically  the
instantonic one by means of extension well-known standard substitutions
(ans\"atzen). In this connection it is possible a reasoning as follows.
All complex-analytic solutions contain real Euclidean ones. On the other
hands, complex-analytic solution is analytic continuation of the Euclidean
solution carrying topological numbers. Can we assert taking into account
holomorphy and continuity that such solutions in pseudoeuclidean space
possess the same topological characteristics?

Other models with internal and gauge symmetries  could be treated within
the approach described. For example, Euclidean $O(2)\times O(2)$-theory
which is widely adopted in different branches of physics, particularly in
the condensed matter physics~\cite{pol}. There are instantons in the model
to be found  by means of $O(2)$-symmetric ansatz. Also the model is quite rich
in duality properties admitting complex structure introduction.

The same is right for $O(3)$-models with curls, monopoles etc.

But of special interest is  the use of these ideas in Gravitation.
Unfortunately, direct attempts to do something like that do not pass trough.
Gravitational gauge potentials in the standard Einsteinian theory
(Cristoffel symbols, riemannian connection) $\Gamma_{\alpha\beta\mu}$ in
coordinate (holonomic) basis are not antisymmetric in $\alpha,\beta$:
$$
\Gamma_{\alpha\beta\mu}=\frac{1}{2}(\partial_\mu g_{\alpha\beta}+
\partial_\beta g_{\alpha\mu}-\partial_\alpha g_{\beta\mu})=
\frac{1}{2}(\partial_\mu g_{\alpha\beta}+\partial_\nu(g_{\alpha\beta\mu}{}^\nu)),
$$
what immediately spoils complexifying, i.e. going over group
$SL(2,C)$. Further, incorporation of the Levi-Civita tensor into
the connection changes geometrical sense of the theory radically.
For example:
$$
\tilde{\Gamma}_{\alpha\beta\mu}=\Gamma_{\alpha\beta\mu}+
\frac{1}{2}\partial_\nu(E_{\alpha\beta\mu}{}^\nu))
$$
Generally speaking, definition of parallel transport  (covariant
derivative) and the curvature tensor form are changed. Unclear,
whether all that can be done  in the consistent way.

Perhaps it would be more logically to treat gravity as a gauge theory.
However there are many other problems there.

Utiyama~\cite{uti} and Kibble~\cite{uti} were the pioneers in  the
development of the approach. Their theories did not coincide  with
General Relativity. In order to obtain the latter one had to
attract ones or others additional considerations. Still so far it
does not  succeed to obtain General Relativity satisfactorily
within the gauge approach~\cite{iva} \footnote{There exist
different viewpoints to the question}. One of the problem here is
presence of tetrad in the formalism together with other field
functions to describe gravity. The tetrad formalism in the {\em
pure} gravity leads to unnecessary dualism in description.

One more argument for {\em pseudoeuclidean} approach rather than Euclidean
one.  In the~\cite{koshk}  method of functional saddle
point (functional steepest descent method) had been proposed. It is shown that within the
framework of some nonperturbation approximation without any going over to
"Euclidean" case the transition amplitude $\int\exp(iS)$ is obtained in the form
of the (nongaussian) path integral of damping exponential.

By this we want to stress that for any arguments {\em pro} 'Euclidean regime' in Physics
one finds their {\em contra}.

\subsection*{Acknowledgment} I am grateful to David Fairlie for his
kindness and discussion.


\end{document}